\documentclass{aa}
\usepackage{natbib}
\usepackage{graphicx}
\def\xmm{{\sc Xmm--newton }}
\def\betam{$\beta$-model } 

\begin{document}

\title{{\sc Xmm--newton} observation of Abell 1835~: temperature, mass and gas mass fraction prof\mbox{}iles.}

\author{S. Majerowicz\inst{1} \and D.M. Neumann\inst{1} \and T.H. Reiprich\inst{2,3}}

\institute{CEA/Saclay, Service d'Astrophysique, L'Orme des Merisiers, B\^{a}t. 709, 91191 Gif--sur--Yvette Cedex, France \and Max--Planck Institut f\"{u}r extraterrestrische Physik, PO Box 1312, 85741 Garching, Germany \and Department of Astronomy, University of Virginia, PO Box 3818, 530 McCormick Road, Charlottesville, VA 22903-0818}

\offprints{S\'{e}bastien Majerowicz,\\ \email{smajerowicz@cea.fr}}

\date{Received  / Accepted }

\abstract{
We present a study of the medium distant cluster of galaxies Abell 1835 based on {\sc Xmm--newton} data. The high quality of \xmm data enable us to perform spectro-imaging of the cluster up to large radii. We determine the gas and total mass prof\mbox{}iles based on the hydrostatic approach using the $\beta$-model and  the temperature prof\mbox{}ile. For the determination of the temperature prof\mbox{}ile of the {\sc icm}, which is needed for the mass determination, we apply a double background subtraction, which accounts for the various kinds of background present (particle and astrophysical background). We f\mbox{}ind a basically f\mbox{}lat temperature prof\mbox{}ile up to 0.75\, $r_{200}$ with a temperature decrease towards the center linked to the cooling f\mbox{}low. We obtain a gas mass fraction of (20.7$\pm 3.7$)\%, which is a lower limit on the baryon fraction in this cluster. Using this value as baryon fraction for the entire universe, we obtain by combining our results with results based on primordial nucleosynthesis, an upper limit for $\Omega_m < 0.5h_{50}^{-1/2}$, which is in good agreement with other recent studies.
}

\authorrunning{S. Majerowicz et al.}
\titlerunning{{\sc Xmm--newton} observation of Abell 1835.}

\maketitle

\section{Introduction}

Clusters of galaxies are dark matter dominated and thus are ideal objects to obtain information on this kind of material. A straightforward way to determine the mass prof\mbox{}ile of a cluster, which consists of roughly 80\% dark matter is to use the hydrostatic approach based on the hot {\sc x}--ray emitting intra-cluster medium (hereafter {\sc icm}). In order to get reliable results with this approach it is necessary that the examined galaxy cluster is in a relaxed state without ongoing merger activity, which might create shock waves and bias the mass estimate. The validity of the hydrostatic approach for relaxed clusters has been tested via hydrodynamic simulations (Schindler 2001~; Evrard, Metzler \& Navarro 1996). To solve the hydrostatic equation one needs to know the density as well as the temperature distribution of the {\sc icm}. The latter one requires {\sc x}--ray observatories, which provide at the same time good spatial as well as spectral resolution. X-ray observatories such as {\sc Rosat}, {\sc Asca}, and {\sc Beppo-Sax} only provided limited combinations of these two requirements and thus the determination of cluster temperature prof\mbox{}iles based on these data were subject to large error bars and sometimes different studies on the same object with the same data came to very different conclusions (Irwin \& Bregman 2000~; Markevitch et al. 1998). A subject of hot debate was the question whether there exists a universal declining temperature prof\mbox{}ile in clusters as suggested by Markevitch et al. (1998) or not. Recently launched {\sc x}--ray satellites namely {\it Chandra} (Weisskopf et al. 2000) and {\sc Xmm--newton} (Jansen et al. 2001) fullf\mbox{}ill now all requirements for an accurate determination of the temperature distribution in clusters up to large radii.  

We present here a spectro-imaging study and subsequent mass determination analysis of the medium distant (z=0.25) relaxed galaxy cluster Abell 1835 (Allen et al. 1992) observed during the {\sc Xmm--newton} Performance Verif\mbox{}ication phase. Abell 1835 is a relaxed cluster, which hosts a cooling f\mbox{}low in its center (Allen et al. 1996). It belongs to the cooling f\mbox{}low clusters for which no multiphase gas down to low temperatures has been found in the {\sc Xmm--newton} {\sc rgs} data (Peterson et al. 2001), and which casts serious doubts on current cooling f\mbox{}low models (Fabian et al. 2001, Molendi \& Pizzolato 2001). 

We determine the total and gas mass prof\mbox{}ile of the cluster up to 1.7\,Mpc which corresponds to 0.75 virial radius (see section \ref{mass}). The ratio of gas mass over total mass allows us to determine a lower limit for the baryon fraction of the cluster which can be considered as representative of the mean baryon fraction of the Universe (see White et al. 1993 and references therein). Knowing the baryon density of the Universe based on studies of primordial nucleosynthesis we can give constraints on the cosmological matter density parameter $\Omega_{\mathrm{m}}$.

Our paper is organized as follows~: in section \ref{observation} we describe the observation. In section \ref{processing} we present the data analysis with specif\mbox{}ic emphasis on vignetting corrections and f\mbox{}lare rejection. In section \ref{data}, we present f\mbox{}irst results and a method to correct for different background components~: a) particle induced background and b) astrophysical background, which varies across the sky (Snowden et al. 1997). The method is based on a double background subtraction. Section \ref{mass} presents our gas and total mass determination and estimate of the baryon fraction of Abell 1835. This is followed in section \ref{discussion} by the discussion of our results and our conclusions in section \ref{conclusions}.

Throughout this article, we use a cosmology with H$_{0}=50$\,km/s/Mpc, $\Omega_{m}=1$ and $\Omega_{\Lambda}=0$ (q$_{0}=0.5$). One arc-minute corresponds to 297 kpc at the cluster redshift of 0.25.

\section{Observation}\label{observation}

{\sc Xmm--newton} observed Abell 1835 in revolution 101 on June 28th 2000. The observation ({\sc id} 98010101) was taken during the phase of Performance Verif\mbox{}ication. The total exposure time was 60\,ksec. Since we are interested in spatially resolved spectroscopy we concentrate here on data from the {\sc epic}-instruments (Turner et al. 2001~; Str\"{u}der et al. 2001). During the observation of Abell 1835 the {\sc mos}1 camera was operating in Large Window mode, which shows a relatively high internal background level, and for which we cannot yet properly account for.  Since our work is very sensitive to background uncertainties we discard {\sc mos}1 data of Abell 1835 in the following and only use data coming from the {\sc mos}2 and pn camera. For all cameras {\sc thin}1 f\mbox{}ilter was used.

\section{Data processing}\label{processing}

We generate calibrated event f\mbox{}iles using the tasks {\em emchain} and {\em epchain} in the {\sc xmm sas} version 5.0 \footnote{\textsf{http://xmm.vilspa.esa.es/sas}}. For the {\sc mos}2 data set, we take into account event patterns 0 to 12 and for pn data only single events ({\sc pattern}=0). 

\subsection{Flare rejection}\label{flare}

\begin{figure}
\resizebox{\hsize}{!}{\rotatebox{-90}{\includegraphics{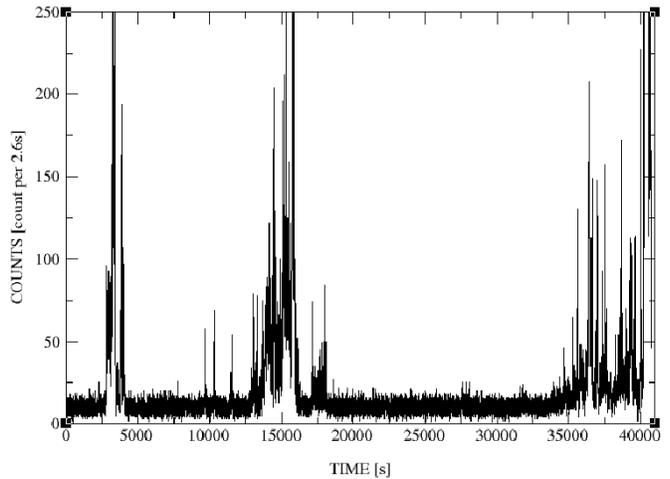}}}
\caption{Light curve derived from {\sc mos}2 camera and event energy between 0.3 and 12\,keV.}\label{lightcurve}
\end{figure}

The light curve of this observation (see f\mbox{}igure \ref{lightcurve}) is not constant and large variations of intensity are visible. These variations, which we will call ``f\mbox{}lares'' in the following, are caused by soft energy protons produced by solar activity, and which are only visible outside the Earth's magnetosphere, which corresponds to the orbit of \xmm (see Ehle et al. 2001).

Unfortunately, the background induced by f\mbox{}lares cannot be easily accounted for since it shows spectral variability. Furthermore, the f\mbox{}lares reduce the observed signal-to-noise ratio. Therefore it is best to discard f\mbox{}lare periods. This reduces the effective exposure time, but improves the signal-to-noise ratio signif\mbox{}icantly.

\begin{figure}
\resizebox{\hsize}{!}{\includegraphics{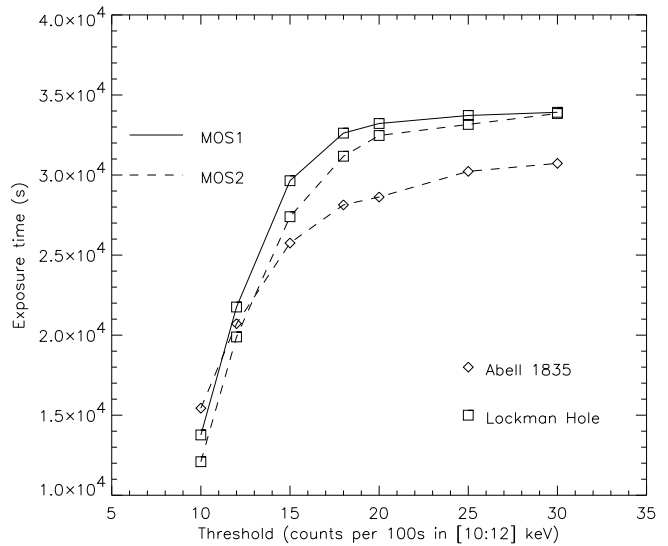}}
\caption{Effective exposure time versus count rate threshold for several {\sc mos} data sets. For the Lockman Hole observations, the revolution 70, respectively 71 was used for {\sc mos}1, respectively {\sc mos}2. The count rate threshold was def\mbox{}ined as the number of events per 100\,s time intervals in the 10 to 12\,keV energy band.}\label{texpomos}
\end{figure}

\begin{figure}
\resizebox{\hsize}{!}{\includegraphics{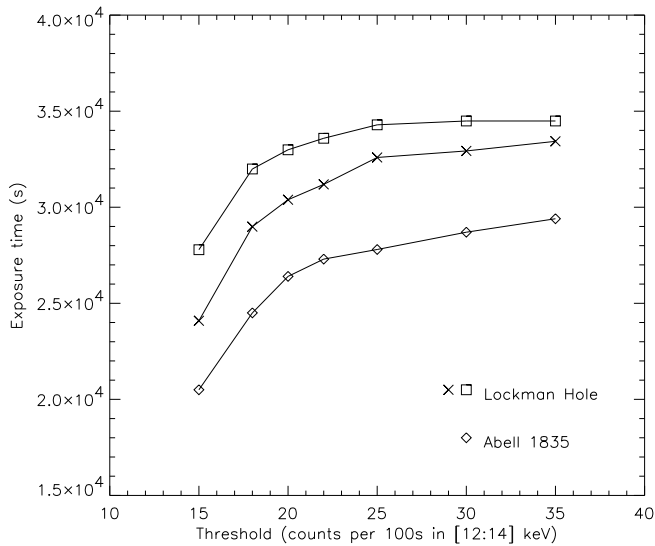}}
\caption{Same as f\mbox{}igure \ref{texpomos} except it is for pn data sets and that the threshold was def\mbox{}ined as the number of events per 100\,s time intervals in the 12 to 14\,keV energy band.}\label{texpopn}
\end{figure}

For our analysis we only consider events inside the f\mbox{}ield-of-view ({\sc fov}) and ignore all events outside the {\sc fov}.

Flares show a hard component observable at high energies easily detectable above 10\,keV where the {\sc mos} cameras are hardly sensitive to X-ray photons. We choose the 10 to 12\,keV energy band to monitor the background for the {\sc mos}2 camera. In order to optimize the background rejection, we analyze the relationship between effective exposure time and the threshold we apply. The results are shown in f\mbox{}igure \ref{texpomos} for Abell 1835 and Lockman Hole (revolution 70 and 71) observations. We can see that the behaviour of exposure time versus rejection threshold is similar for all observations, with a f\mbox{}lattening at a threshold of about 15 counts/100\,s. Thus we can say that this is the optimal threshold for f\mbox{}lare rejection, since it limits exposure time loss.  A higher threshold does not signif\mbox{}icantly increase the exposure time and a lower threshold limits severely the remaining effective exposure time. Therefore we adopt in the following analysis a threshold of 15 counts/100\,s in the 10 to 12\,keV energy band. The same procedure is applied for the pn data (see f\mbox{}igure \ref{texpopn}). The pn camera provides higher quantum eff\mbox{}iciency at high energies than the {\sc mos} detectors. Because of this we shift our energy window for monitoring the background level to the 12 to 14\,keV energy band. Our adopted threshold is 22 counts/100\,s.

Of course, the adopted threshold criterion for f\mbox{}lare rejection must also be applied to the observations chosen to determine the background.

After applying this f\mbox{}lare rejection method, the remaining exposure times are 25750\,s for {\sc mos}2 and 25950\,s for pn. 

\subsection{Vignetting correction}\label{weightmethod}

The collecting area of the {\sc x}--ray telescopes aboard {\sc Xmm--newton} is not constant across the f\mbox{}ield of view. The effective area decreases with increasing off-axis angle and is furthermore photon energy dependent. This effect (vignetting) must be taken into account in the case of extended sources, like galaxy clusters, which cover large fractions of the f\mbox{}ield of view.

To correct for this vignetting effect, we use the method described  by Arnaud et al. (2001a). Brief\mbox{}ly, this method consists of calculating a weight parameter for each event. According to event energy, this parameter is def\mbox{}ined as the ratio of on--axis effective area to effective area at the event position on the detector. Thus, all weight parameters are greater or equal to unity.

\section{Data analysis}\label{data}

\subsection{Morphology}\label{morphology}

\begin{figure}
\begin{center}
\resizebox{\hsize}{!}{\includegraphics[width=6cm]{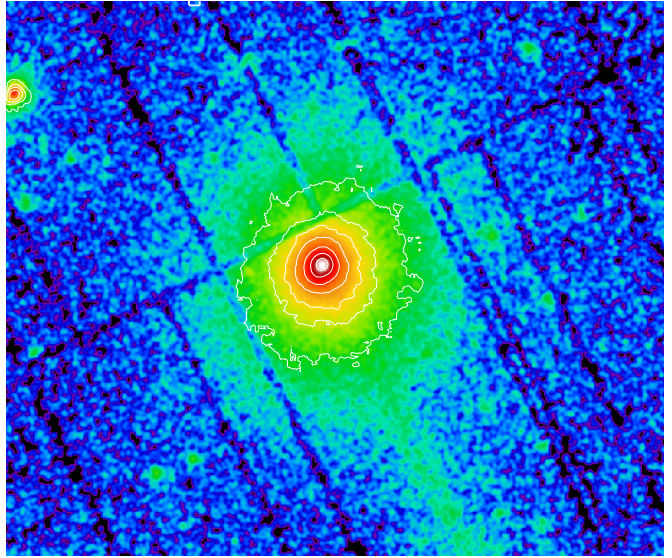}}
\end{center}
\caption{{\sc Mos}2+pn vignetting corrected image of Abell 1835 in sky coordinates in the 0.3 to 3\,keV energy band. This image is f\mbox{}iltered with a $\sigma$=2.2\arcmin\ Gaussian. The contours come from an 0.3 to 3\,keV only {\sc mos}2 image where the gaps between {\sc ccd}s cannot change the shape of these contours. The contour levels are 0.25, 1, 2, 4, 8, 16 and 28 photons by 1.1\arcsec\,$\times$1.1\arcsec\ pixel.}\label{image}
\end{figure}

As can be seen in f\mbox{}igure \ref{image}, Abell 1835 is a compact cluster. There is no strong evidence of substructures which suggests that the cluster is in a relaxed state. The intensity is very peaked in the center due to the presence of a cooling f\mbox{}low (Allen et al. 1996; Schmidt, Allen \& Fabian 2001; Peterson et al. 2001).

\subsection{Spectra}

Since we apply the described vignetting correction method, we can use the on--axis response (i.e. effective area $\times$ redistribution matrix) f\mbox{}iles~: \textsf{m2{\_}thin1v9q20t5r6{\_}all{\_}15.rsp} for the {\sc mos}2 camera and \textsf{epn{\_}ff20{\_}sY9{\_}thin.rsp} for the pn. There exist different response matrix f\mbox{}iles for the pn-chips, which correspond to the different nodes. We use Y9, which is the node farthest away from the readout, with lowest resolution. We tested the different response f\mbox{}iles of the different nodes and could not f\mbox{}ind any detectable difference when performing spectral f\mbox{}its. Due to the remaining uncertainties of the response f\mbox{}iles at very low energy, we exclude events with energy below 0.3\,keV for all spectral f\mbox{}its.

\subsubsection{Background treatment}

Due to the fact that clusters of galaxies are extended and generally low surface brightness sources, the correct determination of the background is important for data f\mbox{}itting and modelling. Since cluster emission decreases with radius from the center, the background component becomes more and more important with increasing radius.

After removing the f\mbox{}lare component by excising high background intensity time intervals, two other background components remain~:
\begin{itemize}
\item high energy particles like cosmic--rays pass through the satellite and deposit a fraction of their energy on the detector. They are not affected by the telescope vignetting. These particles induce instrumental f\mbox{}luorescent lines coming from material aboard the satellite. This kind of background is well monitored by blank--sky observations based on several high galactic latitude observations in which {\sc x}--ray sources were excised. These observations were compiled by D. Lumb (2002)\footnote{They can be retrieved from the Vilspa ftp site~:\ {\small \textsf{ftp://xmm.vilspa.esa.es/pub/ccf/constituents/extras/background}}} for each {\sc epic} camera~;
\item the astrophysical {\sc x}--ray background. The soft component of this background is position dependent and changes across the sky. Therefore it is not necessarily well monitored by blank-f\mbox{}ield observations taken at different positions on the sky.
\end{itemize}

A description of the following method to correct for each background component can also be found in Majerowicz \& Neumann (2001) and Pratt et al. (2001). A similar method was used by Markevitch \& Vikhlinin (2001) for {\em Chandra} data.

In the data sets of Abell 1835 and of the blank-sky observations, which we use for background subtraction, we attribute to each event a weighting factor to correct for vignetting (see section \ref{weightmethod}). This is obviously wrong in case of particle events, which are not vignetted. If the particle background (after f\mbox{}lare rejection, see section \ref{flare}) is constant with time in terms of spectrum shape, what we assume here, we can correct for the falsely attributed weighting factor by subtracting the background spectrum in the same detector region as the source spectrum. In order to account for possible long time intensity variations in the particle induced background, we determine the count rates of source and background observation at high energies ($>10$~keV), where the quantum eff\mbox{}iciency for X-ray photons are low. The ratio of the countrates is used to normalize the corresponding background in the spectral f\mbox{}its. For the {\sc mos}2 we use the 10-12~keV band for determining the background intensity, and for the pn-data  the 12-14~keV band, respectively. In our case, a normalization factor of 1 is found. With this method we correct for the particle background. However we do not yet necessarily account for the spatial variations of the {\sc cxb}.

\begin{figure}
\begin{center}
\resizebox{\hsize}{!}{\includegraphics{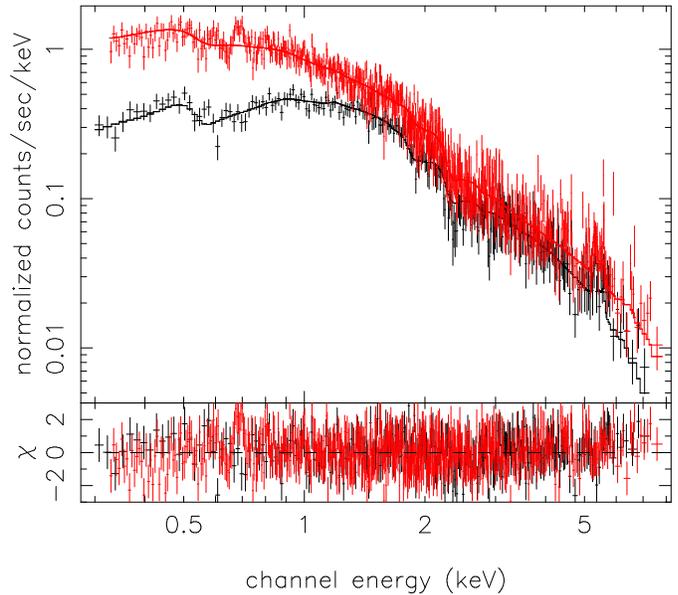}}
\end{center}
\caption{A {\sc mos}2 (lower) and pn (upper) combined f\mbox{}it of fully background corrected spectra in the outside region of the cooling f\mbox{}low (see section \ref{tmean}).}\label{spectre}
\end{figure}

Abell 1835 is located in a peculiar region of the sky, where an excess of low energy {\sc cxb} can be seen (Snowden et al. 1997). To correct for this local excess of the {\sc cxb}, we use regions of the detector in which {\sc x}--ray emission from the cluster is negligible. We assume in the following that the {\sc cxb} does not change signif\mbox{}icantly across the 30\arcmin\ of detector f\mbox{}ield of view. We extract a spectrum in the annulus between 7\arcmin\ and 12\arcmin\ from the cluster center. From this spectrum outside of the cluster, we subtract the blank-sky spectrum extracted from the same detector region. The residual spectrum represents the difference of the {\sc cxb} between the observation of Abell 1835 and the blank-sky observations. These residuals were subsequently subtracted from the cluster spectrum, from which we already subtracted the blank-sky spectrum in the same detector regions. We thus yield spectra which are now properly corrected for the two background components described above. In f\mbox{}igure \ref{spectre}, we present background corrected spectra from {\sc mos}2 and pn data which will be used in  section \ref{tmean} for the estimate of the mean temperature outside the cooling f\mbox{}low region.

For the pn data there exists another source of contamination, the so called out of time ({\sc oot}) events, i.e. events counted during the read-out (see Str\"{u}der et al. 2001). They are visible as a jet--like 
structure in f\mbox{}igure \ref{image}. To estimate the impact of these {\sc oot} events on the temperature determination, we extract spectra in regions def\mbox{}ined in table \ref{tablefit} in which we exclude the {\sc oot} affected regions. Comparing the f\mbox{}it results of those {\sc oot} corrected spectra with the non-corrected {\sc oot} spectra we f\mbox{}ind a slight increase in the temperature estimates of a few percent, however, always well within the error bars. We thus conclude that the {\sc oot}'s do not play an important role in the temperature determination for this particular observation. In the following analysis of pn data, we use the non-{\sc oot} corrected spectra.

\subsubsection{Temperature prof\mbox{}ile}\label{tprofilesection}

\begin{table*}
\begin{center}
\begin{tabular}{cccccc}
\hline
\multicolumn{2}{c}{Annulus (')} & \multicolumn{3}{c}{T (keV) ($\chi^{2}_{\mathrm{red}}$/$dof$)} & A (solar)
\\
$r_{1}$ & $r_{2}$ & {\sc mos}2 & pn & {\sc mos}2+pn & {\sc mos}2+pn
\\
\hline
0.0 & 0.25 & $4.81\pm0.25$ (1.17/307) & $4.43\pm0.12$ (1.17/706) & $4.52\pm0.10$ (1.18/1014) & $0.42\pm0.05$
\\
0.25 & 0.75 & $6.67\pm0.29$ (0.98/371) & $6.16\pm0.20$ (1.12/933) & $6.31\pm0.16$ (1.09/1305) & $0.30\pm0.04$ 
\\
0.75 & 1.5 & $7.29^{+0.45}_{-0.39}$ (0.95/325) & $6.76\pm0.27$ (1.02/749) & $6.94\pm0.22$ (1.00/1075) & $0.23\pm0.05$
\\
1.5 & 2.25 & $7.24^{+0.87}_{-0.63}$ (0.91/229) & $7.15^{+0.50}_{-0.46}$ (1.02/509) & $7.19\pm0.38$ (0.99/739) & $0.32\pm0.08$
\\
2.25 & 3.33 & $6.87^{+1.15}_{-0.95}$ (1.10/164)  & $8.31^{+1.13}_{-1.05}$ (0.90/339) & $7.68^{+0.89}_{-0.73}$ (0.97/504) & $0.27\pm0.13$
\\
3.33 & 6.0 & $6.98^{+3.35}_{-1.89}$ (0.95/53) & $7.80^{+3.90}_{-2.09}$ (1.15/84) & $7.31^{+2.32}_{-1.33}$ (1.06/138) & $<0.9$
\\
\hline
\end{tabular}
\end{center}
\caption{Results for the isothermal model f\mbox{}its derived from equ.\,(\ref{model_fit}) where the free parameters are the temperature T and the abundance A. The galactic hydrogen column density value n$_\mathrm{H}$ is frozen to the galactic value of 2.24$\times$10$^{20}$\,cm$^{-2}$ (Dickey \& Lockman 1990). The spectra are extracted in annuli with inner radius $r_{1}$ and outer radius $r_{2}$. The errors have a conf\mbox{}idence level of 90\%.}\label{tablefit}
\end{table*}

Abell 1835 appears to be a relaxed cluster of galaxies (see section \ref{morphology}). Therefore we assume that the temperature structure of this cluster is spherically symmetric. We extract cluster spectra in concentric annuli around the cluster center. The size we choose for the annuli is a compromise between number of source counts to have suff\mbox{}icient statistics on one hand and small enough regions to determine the temperature prof\mbox{}ile with small bins on the other hand. In our analysis, we exclude the most luminous point sources which are located inside the f\mbox{}ield of view of the cameras.

We group spectra such that each bin has a signal-to-noise ratio greater than 3$\sigma$ after background subtraction. This grouping allows us to assume Gauss statistics, which is important for $\chi^{2}$ f\mbox{}itting.

For the f\mbox{}it to the spectra, we use {\sc xspec} version 11.0.1 (Arnaud 1996) and we model the obtained spectra with~:
\begin{equation}
\mathrm{Model}=\mathrm{Wabs(\mbox{n$_\mathrm{H}$})}\times\mathrm{Mekal(T,A)}\label{model_fit}
\end{equation}
where $\mathrm{Wabs}$ is a photo electric absorption model (Morrisson \& McCammon 1983) and $\mathrm{Mekal}$ a single temperature plasma emission model (Mewe et al. 1985~; Kaastra 1992~; Liedahl et al. 1995). The free parameters are the galactic hydrogen column density value n$_\mathrm{H}$, the {\sc x}--ray temperature kT and the metallicity (abundance) A.

\begin{figure}
\begin{center}
\resizebox{\hsize}{!}{\includegraphics{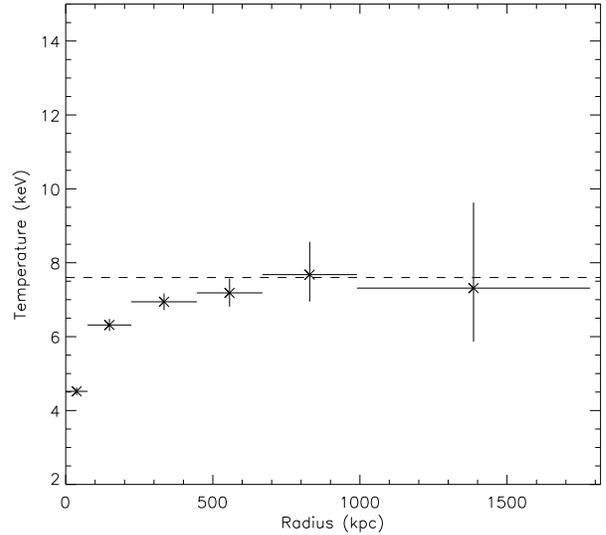}}
\end{center}
\caption{Abell 1835 temperature prof\mbox{}ile (see table \ref{tablefit}) derived from a combined f\mbox{}it of the {\sc mos}2 and pn cameras where the n$_\mathrm{H}$ value is equal to the galactic value at the sky position of Abell 1835. The mean temperature (see section \ref{tmean}) is represented by the dashed line.}\label{tprofile}
\end{figure}

\begin{figure}
\begin{center}
\resizebox{\hsize}{!}{\includegraphics{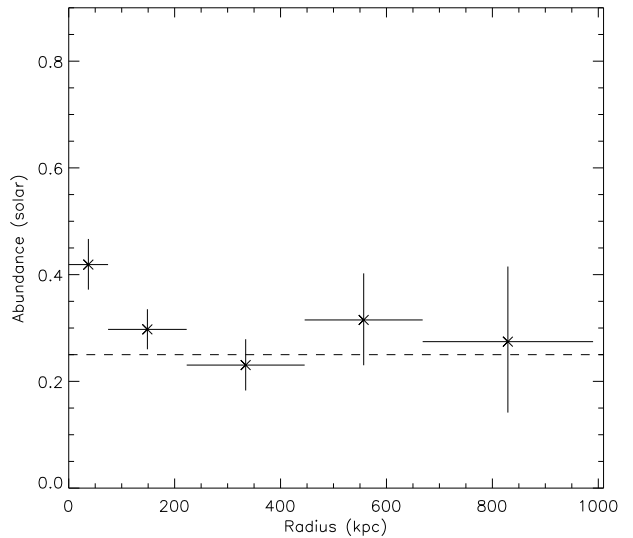}}
\end{center}
\caption{Abundance prof\mbox{}ile (see table \ref{tablefit}) and its mean value outside the cooling f\mbox{}low (dashed line, see section \ref{tmean}) in units of solar metallicity up to 1 Mpc.}\label{aprofile}
\end{figure}

\begin{figure}
\begin{center}
\resizebox{\hsize}{!}{\includegraphics{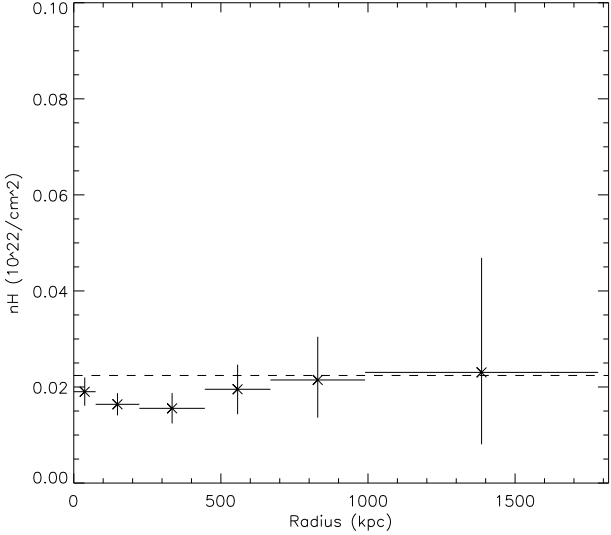}}
\end{center}
\caption{Absorption column density n$_\mathrm{H}$ prof\mbox{}ile from table \ref{tablefitnh}. The dashed line represents the galactic value of 2.24$\times$10$^{20}$\,cm$^{-2}$ (Dickey \& Lockman 1990).}\label{nhprofile}
\end{figure}

First we f\mbox{}it all the spectra with a f\mbox{}ixed galactic absorption n$_\mathrm{H}=2.24\times$10$^{20}$\,cm$^{-2}$ given by Dickey \& Lockman (1990) from 21\,cm line measurements. Results are shown in table \ref{tablefit}. In a second spectral analysis we leave the n$_\mathrm{H}$ value as a free f\mbox{}it parameter (see table \ref{tablefitnh}). The resulting f\mbox{}it temperatures agree very well with those obtained with f\mbox{}ixed n$_\mathrm{H}$ which gives us high conf\mbox{}idence in our spectral analysis and our background subtraction. Figure \ref{nhprofile} shows the f\mbox{}it results for n$_\mathrm{H}$ which are in good agreement with the measurements from the 21\,cm line. Figure \ref{tprofile} shows the resulting temperature prof\mbox{}ile up to a physical radius of 1.8 Mpc derived from the table \ref{tablefit}. Figure \ref{aprofile} shows the metal abundance prof\mbox{}ile up to 1 Mpc.

\begin{table}
\begin{center}
\begin{tabular}{cccc}
\hline
Annulus & T (keV) & n$_\mathrm{H}$ (10$^{20}$\,cm$^{2}$) & $\chi^{2}_{\mathrm{red}}$/$dof$
\\
\hline
1 & $4.60^{+0.23}_{-0.12}$ & $1.90\pm0.30$ & 1.17/1013
\\
2 & $6.61\pm0.20$ & $1.64\pm0.23$ & 1.07/1304
\\
3 & $7.31^{+0.33}_{-0.29}$ & $1.56\pm0.32$ & 0.99/1074
\\
4 & $7.34^{+0.66}_{-0.47}$ & $1.95\pm0.52$ & 0.99/738
\\
5 & $7.58^{+1.13}_{-0.96}$ & $2.16^{+0.90}_{-0.78}$ & 0.97/503
\\
6 & $7.28^{+3.30}_{-1.70}$ & $2.30^{+2.38}_{-1.50}$ & 1.07/137
\\
\hline
\end{tabular}
\end{center}
\caption{Spectral f\mbox{}it results with n$_\mathrm{H}$ value as free f\mbox{}it parameter ({\sc mos}2+pn combined f\mbox{}its).}\label{tablefitnh}
\end{table}

\subsection{{\sc Psf} corrected temperature prof\mbox{}ile}\label{psfsection}

Abell 1835 is a medium distant galaxy cluster with a very peaked surface brightness prof\mbox{}ile (see f\mbox{}igure \ref{sbp}) due to the presence of a cooling f\mbox{}low in the centre. Because of this, the point spread function ({\sc psf}) of {\sc Xmm--newton} (Griff\mbox{}iths et al. 2002, Ghizzardi 2001) potentially blurs the central regions of the cluster, which might inf\mbox{}luence the observed temperature prof\mbox{}ile (see also Markevitch 2002). In order to assess this inf\mbox{}luence we perform another spectral analysis, in which we take into account the {\sc psf} of {\sc Xmm--newton}. We apply the method which is extensively described by Pratt \& Arnaud (2002).

\begin{figure}
\resizebox{\hsize}{!}{\includegraphics{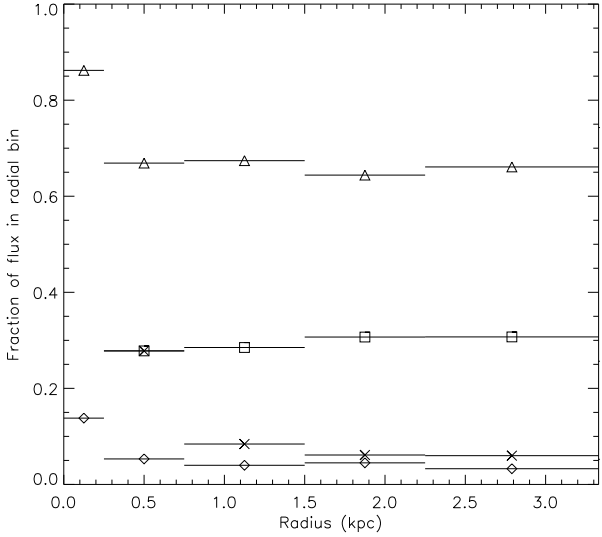}}
\caption{Redistribution of the cluster f\mbox{}lux for the pn camera in the radial bins we use to derive the temperature prof\mbox{}ile. The triangles represents the f\mbox{}lux coming from the bin, the squares from all inner bins to the bin, the losanges from all outer bins to the bin and the crosses from the central bin to the bin.}\label{redipsf}
\end{figure}

For our analysis we assume that the {\sc psf} is constant across the extent of the cluster and that its shape is energy independent. We calculate the f\mbox{}lux redistribution of all temperature bins due to blurring.
This computation is based on a convolution of a double $\beta$-model, which is f\mbox{}itted to the surface brightness prof\mbox{}ile and which gave the following parameters (this f\mbox{}it was performed taking the {\sc psf} into account). Subsequently, based on these model parameters ($\beta_{\mathrm{in}}=0.49$, $r_{c,\mathrm{in}}=0.1$', $\beta_{\mathrm{out}}=0.74$, $r_{c,\mathrm{out}}=0.79$')
we calculate the f\mbox{}lux contribution of bin $i$ to bin $j$. The result can thus be interpreted as a matrix. The corresponding redistribution elements are shown in f\mbox{}igure \ref{redipsf} for the pn data. The redistribution for {\sc mos}2 is similar. As an illustration of the importance of the {\sc psf} effect in the centre~: only 65\% of the f\mbox{}lux of the second bin ($i=2$) are actually observed in the second bin ($j=2$). The remaining 35\% are distributed in other bins ($i=2,j\ne 2$).

\begin{figure}
\resizebox{\hsize}{!}{\includegraphics{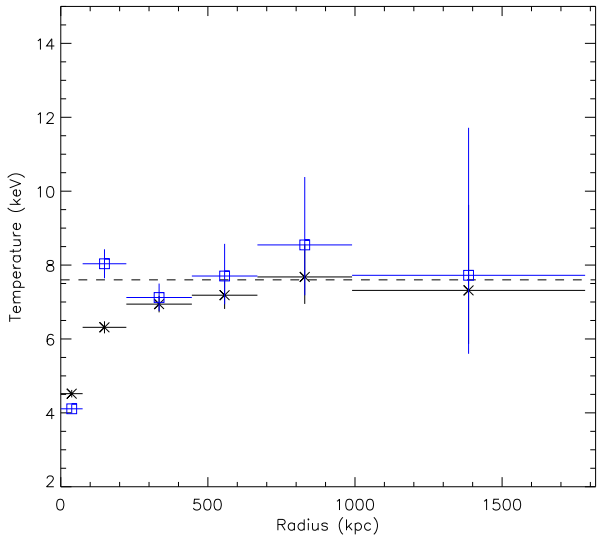}}
\caption{Temperature prof\mbox{}ile of Abell 1835 corrected for the PSF (from table \ref{tablefitpsf}, squares) comparing to the previous one (from table \ref{tablefit}, crosses). The dashed line represents the mean temperature outside the cooling f\mbox{}low region (see section \ref{tmean}).}\label{tprofilepsf}
\end{figure}

After the determination of the redistibution or matrix elements we f\mbox{}it simultaneously the different spectra of our six temperature bins. We take into account the different f\mbox{}lux contributions of each bin by f\mbox{}ixing the ratios of normalization of the f\mbox{}itted spectral models based on our calculated redistribution matrix. Our resulting $\chi^{2}_{\mathrm{red}}$ for the combined f\mbox{}it is $\chi^{2}_{\mathrm{red}}=1.05$ for 4775 degrees of freedom. The resulting temperature prof\mbox{}ile is displayed in f\mbox{}igure \ref{tprofilepsf} as well as in table \ref{tablefitpsf}. Brief\mbox{}ly, when we take into account the {\sc psf} we obtain a steeper temperature gradient in the cooling f\mbox{}low region of the cluster. The temperature estimates at larger radii are hardly affected by the {\sc psf} and are within the error bars of the non-{\sc psf} corrected analysis. Our study presented here is focussed on the external regions of the cluster, therefore we conclude that the {\sc psf} does not have an important impact on our analysis (see below).

\begin{table}
\begin{center}
\begin{tabular}{ccc}
\hline
Annulus & T (keV) & Abundance (solar)
\\
\hline
1 & $4.11\pm0.12$ & $0.43\pm0.05$
\\
2 & $8.03\pm0.39$ & $0.32^{+0.06}_{-0.08}$
\\
3 & $7.12\pm0.36$ & $0.19\pm0.08$
\\
4 & $7.70^{+0.87}_{-0.77}$ & $0.37^{+0.16}_{-0.14}$
\\
5 & $8.55^{+1.36}_{-1.84}$ & $<0.5$
\\
6 & $7.72^{+3.99}_{-2.12}$ & $<1.0$
\\
\hline
\end{tabular}
\end{center}
\caption{Temperature estimate in each annulus for {\sc psf} corrected {\sc mos}2+pn combined f\mbox{}it.}\label{tablefitpsf}
\end{table}

In the following analysis, we will exclusively use the {\sc psf} corrected temperature prof\mbox{}ile.

\subsection{Mean temperature outside the cooling f\mbox{}low}\label{tmean}

We use the background subtracted spectra for {\sc mos}2 and pn data in an annulus between 1\arcmin\ and 6\arcmin\ --- between 300 and 1800\,kpc --- around the cluster center to obtain the mean temperature of the cluster outside the cooling f\mbox{}low region. We perform a combined f\mbox{}it using the model described above in equ.\,(\ref{model_fit}) with a n$_\mathrm{H}$ value f\mbox{}ixed to 2.24$\times$10$^{20}$\,cm$^{-2}$. We f\mbox{}ind a temperature of $7.6\pm0.4$\,keV and an abundance of 0.25$\pm 0.05$ with 90\% conf\mbox{}idence level.

\subsection{Surface brightness prof\mbox{}ile}\label{sbpsection}

\begin{figure}
\resizebox{\hsize}{!}{\includegraphics{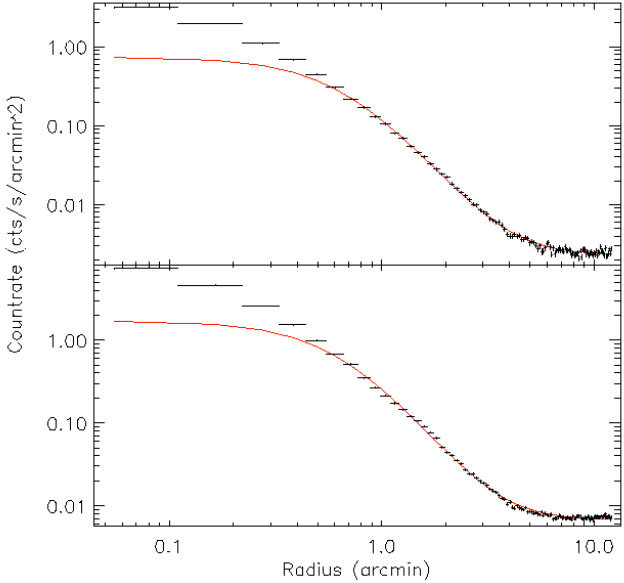}}
\caption{Abell 1835 surface brightness prof\mbox{}iles from {\sc mos}2 (upper panel) and from pn (lower panel) in 0.3 to 3\,keV energy band (1\arcmin corresponds to 297\,kpc). The error bars are one $\sigma$ based on Poisson statistics. The resulting $\beta$-model f\mbox{}its are shown in full line.}\label{sbp}
\end{figure}

We extract surface brightness prof\mbox{}iles from the cluster data. In order to account for vignetting effects we use again the photon weighting method described in section \ref{weightmethod}. In our specif\mbox{}ic case using the weight factor for each event to account for vignetting, the count rate $CR_{i}$ in the i-th annulus is def\mbox{}ined as~:
\begin{equation}
CR_{i}={{1}\over{T_{\mathrm{expo}}}}{{\sum_{j=1}^{n} w_{j}}\over{A_{i}}}
\end{equation}
where $n$ is the number of events in the i-th annulus, $w_{j}$ the event weight parameter, $A_{i}$ the surface of the selected annulus and $T_{\mathrm{expo}}$ the exposure time. In this case, the Poisson error becomes~:
\begin{equation}
\sigma_{i}={{1}\over{T_{\mathrm{expo}}}}{{\sqrt{\sum_{j=1}^{n} w_{j}^{2}}}\over{A_{i}}}
\end{equation}
Serendipitous sources in the f\mbox{}ield of view are excluded. Figure \ref{sbp} shows the surface brightness prof\mbox{}ile of Abell 1835 derived from the {\sc mos}2 and pn cameras in the 0.3 to 3\,keV energy band. 

The surface brightness prof\mbox{}ile is f\mbox{}itted with a $\beta$-model (Cavaliere \& Fusco--Femiano 1976) in which the surface brightness $S$ is def\mbox{}ined as~:
\begin{equation}
S(r)=S_{0}{\Bigg(1+{\bigg({{r}\over{r_{c}}}\bigg)}^{2}\Bigg)}^{-3\beta+{{1}\over{2}}}+B
\end{equation}
where $S_{0}$ is the central intensity, $r_{c}$ the core radius, $\beta$ the slope parameter and $B$ the background intensity. All these parameters must be f\mbox{}itted.

Abell 1835 hosts a cooling f\mbox{}low in its center (see above). Figure \ref{sbp} shows the characteristic peak of intensity for cooling f\mbox{}lows towards the center (e.g. Fabian 1994). In order to avoid f\mbox{}itting in the cooling f\mbox{}low region we choose to exclude a region with a radius of roughly 0.1\,r$_{200}$ (see equ.\,(\ref{r200})) --- i.e. 0.7\arcmin --- around the center for the $\beta$-model f\mbox{}it. Indeed Neumann \& Arnaud (1999) showed that below 0.1\,r$_{200}$ the surface brightness prof\mbox{}iles for nearby clusters show a large dispersion due to cooling f\mbox{}low but they look remarkably similar above  0.1\,r$_{200}$. The best f\mbox{}it parameters are $\beta=0.704\pm 0.005$ and a core radius $r_{c}=202.3\pm 7.1$\,kpc (i.e. $r_{c}=0.681\pm 0.024$\arcmin ) for the {\sc mos}2-data. The uncertainties are 68\% conf\mbox{}idence level. We also perform a $\beta$-model f\mbox{}it with the surface brightness prof\mbox{}ile extracted from the pn data for consistency. We f\mbox{}ind in this second case $\beta=0.708^{+0.012}_{-0.007}$, $r_{c}=196.9^{+5.1}_{-5.3}$\,kpc (i.e. $r_{c}=0.663^{+0.017}_{-0.018}$\,\arcmin). These f\mbox{}it parameters are thus in good agreement with those found with the {\sc mos}2 camera.

Our found parameters of $\beta$ and $r_c$ are in relatively good agreement with the results found by Neumann \& Arnaud (1999) for nearby clusters.

\section{Mass analysis}\label{mass}

We assume in the following that the cluster is virialized up to $r_{200}$, the radius in which the mean density of the cluster is 200 times the critical mass density. For the calculation of $r_{200}=r_{\rm{virial}}$ we use the relation by Evrard et al.
(1996)~:
\begin{equation}
r_{200}=3690\sqrt{{{T}\over{10\,\mathrm{keV}}}}{(1+z)}^{-{{3}\over{2}}}\label{r200}
\end{equation}
Using the mean temperature of 7.6 keV (see section \ref{tmean}), we f\mbox{}ind $r_{200}= 2293$\,kpc.

\subsection{Intra cluster medium mass}

Assuming that the radial {\sc icm} distribution follows a $\beta$-model, the electron density $n_{e}$ can be written~:
\begin{equation}
n_{e}(r)=n_{e0}{\Bigg(1+{\bigg({{r}\over{r_{c}}}\bigg)}^{2}\Bigg)}^{-{{3}\over{2}}\beta}\label{ne}
\end{equation}
From the electron density we calculate the total gas density of the {\sc icm} and determine the {\sc icm} mass by integrating equ.\,(\ref{ne}). We obtain a central electron density $n_{e0}=1.47\pm 0.02 \times 10^{-2}$\,cm$^{-3}$ (68\% conf\mbox{}idence level) assuming a temperature of 7.6\,keV and using the $\beta$-model parameters determined in section \ref{sbpsection}. Temperature variations in the uncertainties do not change our values for the electron density.  The error bars of $n_{e0}$ are calculated from the errors of the $\beta$-model parameters. We verify that the cluster temperature has no effect on the estimate of $n_{e0}$ with the selected energy range. 

\begin{figure}
\begin{center}
\resizebox{\hsize}{!}{\includegraphics{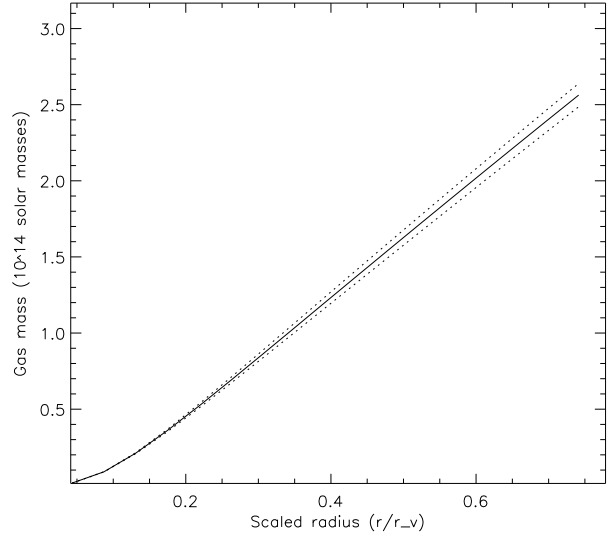}}
\end{center}
\caption{Abell 1835 gas mass prof\mbox{}ile. The error bars (dotted line) are a conf\mbox{}idence level of 68\%. The gas mass prof\mbox{}ile (full line) is derived from the best f\mbox{}it parameters.}\label{gasmass}
\end{figure}

The {\sc icm} mass prof\mbox{}ile is shown in f\mbox{}igure \ref{gasmass}. We f\mbox{}ind M$_{\mathrm{gas}}=2.56\pm 0.08\times 10^{14}$\,M$_{\odot}$ within $r=1.7$\,Mpc or 0.75\,$r_{200}$ with a conf\mbox{}idence level of 68\%.

We want to stress that we underestimate the central gas density in the centre due to neglecting the cooling f\mbox{}low excess. However, this underestimation at small radii does not play a signif\mbox{}icant role when we integrate the gas mass out to large radii.

In fact, when we f\mbox{}it the surface brightness prof\mbox{}ile (f\mbox{}igure \ref{sbp}) so that we only account for the inner 210\,kpc, we f\mbox{}ind $\beta$=0.49, $r_{c}$=52\,kpc and $n_{e0}=5.55\times 10^{-2}$\,cm$^{-3}$. By integrating equ.\,(\ref{ne}) with these parameters, we f\mbox{}ind a gas mass at r=0.1\,$r_{200}$, which is 18\% higher than the one using the global \betam f\mbox{}it parameters. At $r=0.75 r_{200}$, which is the radius up to which we can constrain the temperature and thus total mass prof\mbox{}ile we underestimate the gas mass by 1\% due to neglecting the gas mass linked to the central gas excess with respect to our global \betam, which is negligible.

\subsection{Cluster total mass}

Assuming spherical symmetry and hydrostatic equilibrium, we calculate the gravitational mass of the cluster Abell 1835 with the hydrostatic equation~:
\begin{equation}
\mathrm{M}_{\mathrm{tot}}(<r)=-{{k\mathrm{T}}\over{G\mu m_{p}}}r\Bigg({{d\ln n_{e}}\over{d\ln r}}+{{d\ln \mathrm{T}}\over{d\ln r}}\Bigg)\label{totalmassequ}
\end{equation}
where $k$ is the Boltzmann constant, T the gas temperature, $G$ the gravitational constant, $\mu$ the mean molecular weight of the gas ($\mu\sim 0.6$), $m_{p}$ the proton mass and $n_{e}$ the electronic density. Including equ.\,(\ref{ne}) in equ.\,(\ref{totalmassequ}), M can be written~:
\begin{equation}
\mathrm{M}_{\mathrm{tot}}(<r)=-{{k}\over{G\mu m_{p}}}r^{2}\Bigg({{d\mathrm{T}}\over{dr}}-3\beta \mathrm{T}{{r}\over{r^{2}+r_{c}^2}}\Bigg)\label{mtotbeta}
\end{equation}

\begin{figure}
\begin{center}
\resizebox{\hsize}{!}{\includegraphics{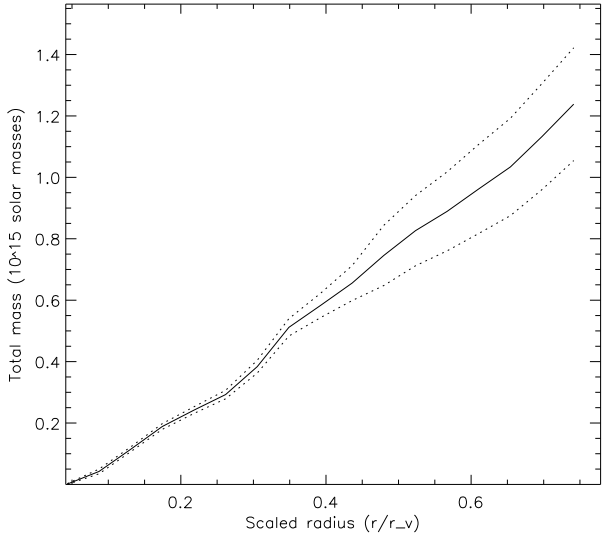}}
\end{center}
\caption{Abell 1835 total mass prof\mbox{}ile using the {\sc psf} corrected temperature prof\mbox{}ile (see section \ref{psfsection}). Error bars are at 68\% conf\mbox{}idence 
level(same display convention as f\mbox{}igure \ref{gasmass}).}\label{totalmass}
\end{figure}

In order to account for the error bars in the temperature prof\mbox{}ile, we use a Monte--Carlo method described by Neumann \& B\"{o}hringer (1995). This method allows to transform error bars of the temperature prof\mbox{}ile into error bars of the mass prof\mbox{}ile. This transformation is done by calculating randomly temperature prof\mbox{}iles, which f\mbox{}it in the actual measured temperature prof\mbox{}ile and to subsequently calculate the mass prof\mbox{}ile corresponding to the randomly determined temperature prof\mbox{}ile. We determine 10000 random temperature prof\mbox{}iles. From the corresponding 10000 mass prof\mbox{}iles we calculate the mean mass prof\mbox{}ile and the errors. The uncertainties are determined by taking into account the uncertainties of the temperature prof\mbox{}ile (90\% conf\mbox{}idence level) and the standard deviation of the mass at a given radius. The resulting errors are calculated so that they correspond to 1$\sigma$ error bars (68\% conf\mbox{}idence level). The mass prof\mbox{}ile is shown in Fig.\ref{totalmass}. At a radius corresponding to 75\% of the virial radius (or 1.7\,Mpc) we obtain a total mass\footnote{For a comparison, the total mass is $1.14\pm0.14\times 10^{15}$\,M$_{\odot}$ when we use the temperature prof\mbox{}ile without {\sc psf} correction (table \ref{tablefit}).} of~: $M_{tot}=1.23\pm0.18\times 10^{15}$\,M$_{\odot}$ (68\% conf\mbox{}idence level).

Assuming that Abell 1835 is isothermal with a temperature of 7.6\,keV (see section \ref{tmean}), equ.\,(\ref{mtotbeta}) can be written as~:
\begin{equation}
\mathrm{M}_{\mathrm{tot}}(<r)={{3k\beta}\over{G\mu m_{p}}}\mathrm{T}{{r^{3}}\over{r^{2}+r_{c}^2}}
\end{equation}
In this case taking into account error bars on T, $\beta$ and $r_{c}$ parameters, the total mass of Abell 1835 is $1.00\pm 0.06\times 10^{15}$\,M$_{\odot}$ which is consistent with the mass obtained using the Monte-Carlo approach. This was expected since the temperature prof\mbox{}ile of Abell 1835 (see f\mbox{}igure \ref{tprofilepsf}) seems to be isothermal at high radii.

\subsection{Gas mass fraction}

\begin{figure}
\begin{center}
\resizebox{\hsize}{!}{\includegraphics{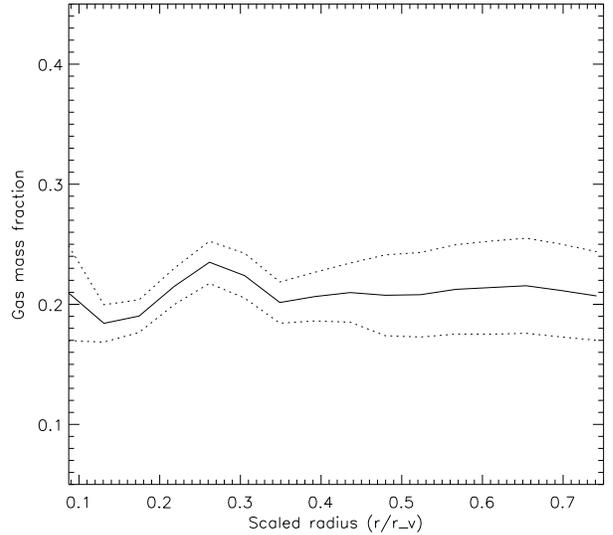}}
\end{center}
\caption{Abell 1835 gas mass fraction prof\mbox{}ile. Error bars are at 68\% 
conf\mbox{}idence level (same display convention as f\mbox{}igure \ref{gasmass}).}\label{fgas}
\end{figure}

The gas mass fraction is simply the ratio of the {\sc icm} mass to the total mass. The prof\mbox{}ile of the gas mass fraction of Abell 1835 is shown in f\mbox{}igure \ref{fgas}. We do not display the gas mass fraction below 10\% of the virial radius because of the cooling f\mbox{}low region and our underestimate of the gas mass there (see section \ref{sbpsection}). The errors bar are calculated with classical error propagation and are at a conf\mbox{}idence level of 68\%. 
Within $r=0.75\,r_{200}$, we f\mbox{}ind $f_{gas}h_{50}^{-1.5}=0.207\pm 0.037$ where $h_{50}$ is the present time Hubble constant in units of 50\,km/s/Mpc.

\section{Discussion}\label{discussion}

\subsection{The gas mass fraction and implications for cosmology}

Abell 1835 is an ideal cluster to determine the gas mass prof\mbox{}ile with the hydrostatic approach, since the {\sc Xmm--newton} image does not show indications of substructure at large scales\footnote{Schmidt et al. (2001) reported small scale substructure in the {\it Chandra} data in the cooling f\mbox{}low region, which does not inf\mbox{}luence our results on large scales.}, which might contaminate the mass determination.

Our determined gas mass fraction prof\mbox{}ile of Abell 1835, which we determine up to 0.75\,$r_{200}$ is essentially f\mbox{}lat. The mean value is about 21\% and within 0.75\,$r_{200}$ we obtain a gas mass fraction of $0.207\pm 0.037 h_{50}^{-3/2}$. We can take this gas mass fraction as lower limit for the baryon fraction in the cluster because we do not take into account other contributions of baryons, such as stars or interstellar medium. Extrapolating this lower limit through out the universe (which was shown to be valid by White et al. (1993) and references therein), we can determine the cosmological matter density parameter $\Omega_m$ by combining our result with studies based on primordial nucleosynthesis, which give f\mbox{}irm constraints on the density parameter of baryons in the universe. Recent results give $\Omega_b h_{50}^2 = 0.080\pm0.008$ (Burles et al. 2001). We can thus apply~:
\begin{equation}
{{M_{\mathrm{gas}}}\over{M_{\mathrm{tot}}}} = 0.207\pm 0.037 h_{50}^{-3/2} = {{\Omega_{\mathrm{b}}}\over{\Omega_{\mathrm{m}}}} = {{0.08\pm0.008 h_{50}^{-2}}\over{\Omega_{\mathrm{m}}}}\label{cosmoequ}
\end{equation}
If the {\sc icm} was the only source of baryons in clusters, above equation gives us $\Omega_{\mathrm{m}}=0.386\pm0.108$ (68\% conf\mbox{}idence level). However, since there exist other baryon sources in Abell 1835 such as galaxies, this implies that our result is an upper limit with~: $\Omega_{\mathrm{m}} < 0.5 h_{50}^{-1/2}$. This result is in very good agreement with other recent results based on {\sc cmb} measurements combined with redshift surveys (Efstathiou et al. 2001) and studies on distant supernov{\ae} (Perlmutter et al. 1999) which favor a low density universe with a cosmological constant. If we additionally knew the baryon fraction in galaxies up to high precision, we could put even tighter constraints on $\Omega_m$.

\subsection{The temperature prof\mbox{}ile -- comparison with other recent studies}

A lot of effort has been spent already for the determination of temperature prof\mbox{}iles in the {\sc icm}. The corresponding results vary from each study to another quite signif\mbox{}icantly. There have been reported declining temperature prof\mbox{}iles, with a more or less universal shape based on {\sc Asca} and {\sc Beppo-sax} data (Markevitch et al. 1998, de Grandi \& Molendi 2001). Other studies based on data from the same instruments, on the contrary, f\mbox{}ind f\mbox{}lat temperature prof\mbox{}iles or prof\mbox{}iles with large dispersions (Irwin \& Bregman 2000, White et al 2000).

All these studies are limited to radii smaller than 0.5\,$r_{200}$ due to instrumental limitations. Furthermore, studies based on {\sc Asca} and {\sc Beppo-sax} were subject to possible systematic uncertainties due to the large point spread function of the mirrors, which additionally can vary in size depending on energy. {\it Chandra} results presented by Allen, Schmidt \& Fabian (2001) suggest that massive, relaxed clusters are isothermal up to $r_{2500}$ and show a universal temperature decline in the center linked to the cooling f\mbox{}low.

{\sc Xmm--newton} with its better spatial resolution than {\sc Asca} and {\sc Beppo-sax} and larger effective area than comparable observatories allows to a) better constrain the temperature prof\mbox{}iles and b) to trace them out to larger radii. So far studies based on {\sc Xmm--newton} show either f\mbox{}lat prof\mbox{}iles (Arnaud et al. 2001a~; Tamura et al. 2001) for hot clusters, which is in agreement with the results by Allen et al. (2001) on {\it Chandra} data of clusters. A \xmm study on the cool cluster S\'{e}rsic 159-03, however, shows a dramatic temperature decline in the outer parts of the cluster (Kaastra et al. 2001). 

\begin{figure}
\begin{center}
\resizebox{\hsize}{!}{\includegraphics{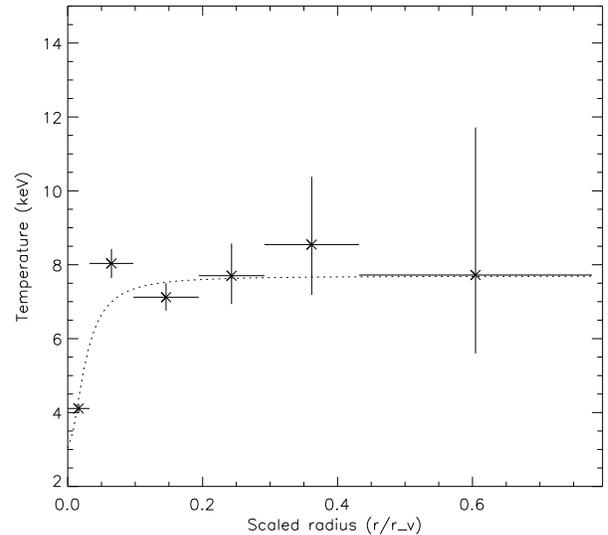}}
\end{center}
\caption{Abell 1835 temperature prof\mbox{}ile and the comparison with the universal prof\mbox{}ile derived by Allen et al. (2001) and extrapolated to 0.75\,$r_{200}$ (dotted line)}\label{tprofilecomp}
\end{figure}

Our results suggest strongly a f\mbox{}lat temperature prof\mbox{}ile with a decline towards the center linked to the cooling f\mbox{}low present in Abell 1835 (Peterson et al. 2001, Schmidt et al. 2001 and references therein). This prof\mbox{}ile f\mbox{}its well with the empirically found model by Allen et al. (2001), which is valid up to $r_{2500}$, which corresponds roughly to $0.3\times r_{200}$ and which is displayed in Fig \ref{tprofilecomp}. However, an individual study undertaken by Schmidt et al. (2001) on Abell 1835 observed with {\it Chandra} shows a temperature prof\mbox{}ile which is increasing with radius. At radii larger than 400\,kpc Schmidt et al. (2001) f\mbox{}ind temperature estimates of about 12\,keV, which is inconsistent with our {\sc Xmm--newton} results. The fact that we f\mbox{}ind good agreement between the individual spectral modeling of {\sc mos}2 and pn cameras gives us conf\mbox{}idence in our results.

While undertaking our study concerning the background subtraction we observed that minor differences in the extraction of f\mbox{}lares between background and source observation can lead to declining or increasing temperature prof\mbox{}iles. Doing monitoring on the background level we f\mbox{}ind that quiescent periods do not always have the same background level and have sometimes, even though rarely, intensities close or above the threshold criterion to extract f\mbox{}lares. In order to screen the data properly for background variations and to determine the level of the particle (instrumental) background accurately it is important to look at parts of the observation where no astrophysical source emission can be detected like the high energy part of the spectrum, where only particle background can be seen (the particle background has a hard spectrum and the effective area of {\sc Xmm--newton} at energies above 10\,keV for {\sc x}--ray photons is relatively small). It is important to compare and to apply exactly the same f\mbox{}lare rejection for both the source and the background observation to avoid wrong background subtraction, which can bias the temperature estimates. Observing a temperature prof\mbox{}ile which increases with radius can be an indication of a slightly higher particle background in the source observation than in the background observation.

\begin{figure}
\resizebox{\hsize}{!}{\includegraphics{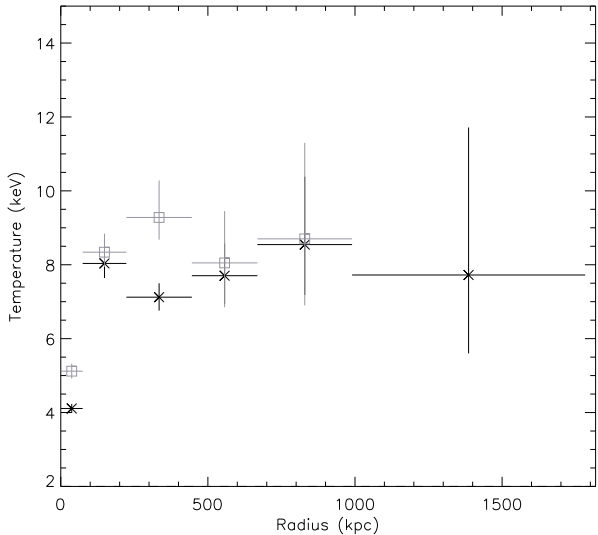}}
\caption{Temperature prof\mbox{}ile of Abell 1835 corrected for the PSF (from table \ref{tablefitpsf}, crosses) comparing to the {\em Chandra} one (squares) obtained by Markevitch (2002).}\label{tprofilechandra}
\end{figure}

The presence of remaining solar f\mbox{}lares in the analysis of Schmidt et al. (2001) was recently discussed by Markevitch (2002), who reanalyzed the {\em Chandra} data of Abell 1835. Markevitch (2002) found that a temperature prof\mbox{}ile analysis based on a more thorough f\mbox{}lare treatment leads to a basically f\mbox{}lat temperature prof\mbox{}ile in the outer parts of the cluster, which agrees well with our derived temperature prof\mbox{}ile (see f\mbox{}igure \ref{tprofilechandra}) at large scales.

\subsection{Global temperature estimates}

There is some dispersion observable between past temperature determinations of Abell 1835 based on {\sc Asca}-data. Allen \& Fabian (1998) found a temperature of $9.8^{+2.3}_{-1.3}$\,keV, based on {\sc Asca-sis} and {\sc gis}-data, including a multiphase cooling f\mbox{}low model, which is now known not to be an accurate description for the cooling f\mbox{}low of Abell 1835 (Peterson et al. 2001). White (2000) only used {\sc Asca-gis} data and obtained an overall temperature of kT=$8.5^{+1.5}_{-0.5}$\,keV when applying a multiphase cooling f\mbox{}low model and kT=$8.2\pm0.5$\,keV when using a single temperature model.  Both results are consistent within the error bars with our results of kT=7.6$\pm 0.4$\,keV. Furthermore, Mushotzky \& Scharf (1997) found a temperature estimate of kT=$8.2\pm0.5$\,keV, which is again comparable with our results and the ones of White (2000). The results of Allen \& Fabian (1998) agree within the error bars with other results based on {\sc Asca}, but are higher than our mean temperature. While our temperature results are somewhat lower than the results of {\sc Asca}, the {\it Chandra} results for the temperature of Abell 1835 by Schmidt et al. (2001) are above the results based on {\sc Asca}.

\section{Conclusion}\label{conclusions}

\xmm data allow us to measure the temperature prof\mbox{}ile of Abell 1835 up to 0.75\,$r_{200}$. In order to determine the temperature prof\mbox{}ile of the cluster accurately we apply a method, which corrects for the various kinds of background and a photon weighting method which allows to correct for vignetting effects. We correct the temperature prof\mbox{}ile for {\sc psf}-effects and see that the resulting prof\mbox{}ile is not affected by the {\sc psf} at large radii.

Our resulting {\sc psf} corrected temperature prof\mbox{}ile is consistent with being f\mbox{}lat in the outer parts. The mean cluster temperature is  7.6\,keV outside the cooling f\mbox{}low region. In the central parts, below a radius of 1\arcmin (300\,kpc) we see a temperature decline linked to the cooling f\mbox{}low of the cluster. We f\mbox{}it a \betam to the outer radii of the surface brightness prof\mbox{}ile of Abell 1835 and obtain the best f\mbox{}it results of $\beta=0.704\pm0.005$ and $r_{c}=202.3\pm7.1$\,kpc. We use these best f\mbox{}it parameters as input for the hydrostatic approach to calculate the mass prof\mbox{}ile of the cluster and f\mbox{}ind a mass at 0.75\,$r_{200}$ of $1.23\pm0.18\times 10^{15}$\,M$_{\odot}$. The corresponding gas mass fraction, which is constant with radius at radii $r>0.1\,r_{200}$, is with 20.7\%$\pm3.7$ comparable with the results found in other studies. 

Our global temperature estimate for Abell 1835 based on \xmm is somewhat low when compared to other results, but agree in the error bars with other studies based on {\sc Asca}-data. We assume that these discrepencies are linked to small remaining uncertainties in the calibration of \xmm , and on which is still being worked on.

\begin{acknowledgements}

We would like to thank Monique Arnaud for useful discussions, suggestions on the background subtraction, and a very important participation in the development of the codes we use here. Equally we are greatful to Ren\'{e} Gastaud, who wrote most of the codes concerning the vignetting correction. Furthermore we would like to thank D. Lumb for his effort to accumulate blank-sky observations, which is essential for our study presented here. We also thank the referee for useful remarks. T.~H.~R. thanks the Celerity Foundation for support. T.~H.~R. was also supported by NASA XMM-Newton Grant NAG5-10075. Also, we are very grateful to the entire {\sc epic}-calibration team, which shared all essential information with us, which made this study possible. The results presented here are based on observations obtained with {\sc Xmm--newton}, an {\sc Esa} science mission with intruments and contributions directly funded by {\sc Esa} Member States and the {\sc Usa} ({\sc Nasa}).

\end{acknowledgements}

\end{document}